\documentclass[%
 reprint,
 amsmath,amssymb,
 aps,
]{revtex4-2}

\usepackage{graphicx}
\usepackage{dcolumn}
\usepackage{bm}

\usepackage{lineno}
\usepackage{prettyref}
\usepackage{xcolor}

\usepackage{hyperref}

\begin{document}

\title{On the neutron orbital angular momentum: 
Has it  actually been demonstrated? }
\author{Wolfgang Treimer}
	  \email{wolfgang.treimer@bht-berlin.de}
\author{Frank Haußer}
	 \email{frank.hausser@bht-berlin.de}
	\affiliation{Department of Mathematics, Physics and Chemistry, Berliner Hochschule für Technik, D-13353 Berlin, Germany
}
\author{Martin Suda}
\email{Martin.Suda@ait.ac.at}
\affiliation{
Security $\&$ Communication Technologies, Center for Digital Safety $\&$  Security,
AIT  Austrian Institute of Technology GmbH, A-1210 Vienna, Austria
}
%
%

\date{\today}

\begin{abstract}

{{In a Nature LETTER \cite{Clark}, a control of neutron angular momentum was apparently demonstrated using a neutron crystal  interferometer. In the meantime, a number of highly interesting articles have been published dealing with the n-OAM and neutron interferometry, citing \cite{Clark}.  We show that the interpretation of \cite{Clark} having detected a n-OAM is incorrect for several serious reasons and point out  inconsistencies. Based only on the theory of dynamical neutron diffraction, we calculate all the interferograms and show perfect agreement with those given in \cite{Clark}. There was no need to assume an additional n-OAM because no additional effect was observed that could not be explained by standard theory.
}
}
\end{abstract}

\keywords{Neutron orbital angular momentum, neutron interferometry}
\maketitle
\section{}

\section{Introduction} \label{Intro}

The control of the orbital angular momentum (OAM) of neutrons is expected to provide additional quantum information and new aspects for the study of the foundations of quantum mechanics \cite{Clark}. 
Quantized orbital angular momentum has already been realized for photons and used, for example, for quantum entanglement, quantum information science, and imaging \cite{Mair2001}, \cite{Verbeeck2010}.  The use of electron orbital angular momentum  offers exciting applications \cite{McMorran2011}, \cite{Krenn2017}. 
 {{Recently X-rays with orbital angular momentum could be generated in a free-electron laser oscillator \cite{Nanshun2021}, twisting neutral particles with electric fields  and phase vortex lattices in neutron interferometry were investigated \cite{Geerits2021}, \cite{Geerits2023}. Theoretical work on spin-textured neutron beams with orbital angular momentum all referring \cite{Clark} was recently published \cite{Pynn2023}.}}
A comprehensive review of optical vortices, i.e.  OAM manipulations from topological charge to multiple singularities, can be found in \cite{Rubinsztein2017},  \cite{Shen2019}.  
Therefore,  a use of a neutron  orbital angular momentum (n-OAM)  is  extremely interesting \cite{Sarenac2016,Sarenac2018, Sarenac2019,Jiang2020}. 
However, the evidence of the measurement of a controlled n-OAM given in \cite{Clark}  is highly implausible and leads to severe, far reaching  misinterpretations,  once due to the lack of sufficiently large coherence of the neutron beam \cite{Cappelletti2018,Jach2022}, then due to other much more severe reasons, given below. 
We provide cogent explanations why the detection of n-OAM in \cite{Clark} is unlikely, clarify the conditions for this detection,  and calculate  all interferograms given in  \cite{Clark}, by using exactly the same parameters, i.e.  perfect crystal  neutron interferometer, neutron wave length, spiral phase plate  and neutron detector, using only the theory of dynamical neutron diffraction. 
Our results are in perfect agreement with those presented in \cite{Clark}.

The first evidence that light can have angular orbital  momentum was published as early as 1935, followed by extensive studies on this phenomenon many years later \cite{Beth1935,Beth1936,
Allen1992,Barnett1994}. It was found that light with a spatially varying amplitude and phase distribution due to an object with a spiral phase contains an OAM  and that this should be quantized in units of $ \hbar$. Therefore, it seemed logical to generate n-OAM  with a spiral phase plate and to use a neutron interferometer to detect it.  

\section{Neutron crystal interferometry}

{All mathematics and theory  concerning neutron interferometry are given in \cite{RauchWerner2000}, however,  some details need to be explained here because they are not so well known, but are important to understand interferograms produced by a crystal interferometer.}
\begin{figure}[h]
\centering
\includegraphics[width=\columnwidth]{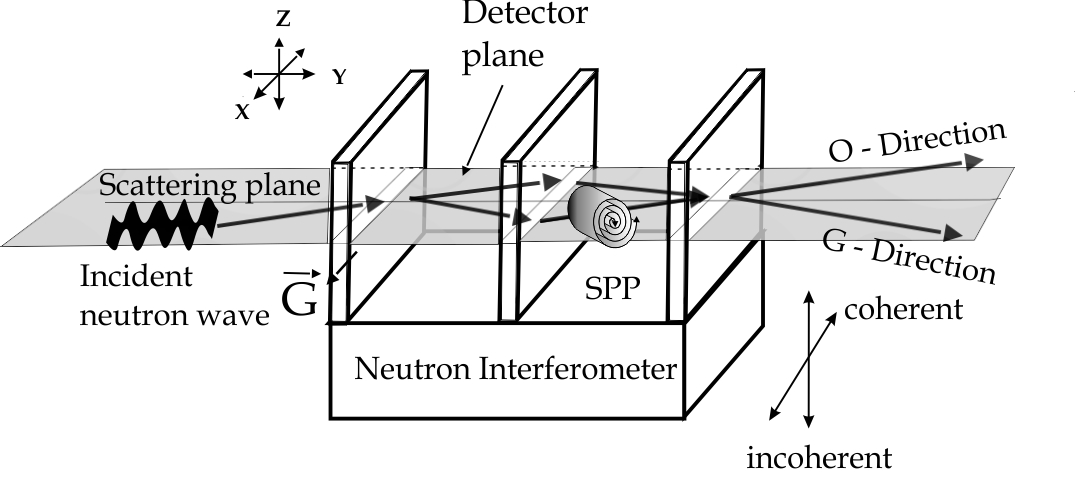} 
\caption{\textbf{Layout of a  neutron interferometer}  $|$  
By the first crystal plate, the incident neutron wave is coherently split into a forward beam (O-direction) and a Bragg diffracted  beam (G-direction), which is repeated by the second and third crystal plate. Only the interfering beams are depicted. 
Note, it makes no difference whether the spiral phase plate (SPP) is in the other beam path. The gray area is the scattering plane, defined by the  incident wave vector  of the neutron and the reciprocal lattice vector  $ \vec{G} $. }
\label{Fig_1}
\end{figure}

{{ In order to use a neutron interferometer (Fig.~\ref{Fig_1}) for a specific purpose, it is necessary to know that it is a crystal interferometer and that the underlying interaction of the neutron waves is crystal diffraction.  } }
For  X-rays this is given by the dynamic theory of diffraction (\cite{Laue1960,Zachariasen1945}),  which later on was adapted to  neutron interferometry  (for details see  \cite{RauchWerner2000,Petrascheck1988}).  
The fundamental idea of perfect crystal interferometry is  coherent splitting of a  beam by so-called Laue-diffraction (first interferometer crystal plate, splitter), {which is   repeated}  by a second (mirror crystal) and  the  third crystal plate (analyzer crystal) as shown in Fig.~\ref{Fig_1}.  
The coherent superposition of two converging beams in front of  or inside  the analyzer crystal produces an interference pattern  whose fringe spacing  {d$ _{f} $}  is equal to the crystal lattice spacing  (for  $\lambda $ = 0.271nm  {d$ _{f} $}   $ \sim $ 0.314 nm). This interference pattern cannot be resolved by any X-ray or neutron detector. 
Instead, the reflecting atomic planes of the analyzer crystal are used to create a Moiré pattern with a fringe pattern magnified by about 10$ ^{7} $, which makes the observation of neutron interferences possible at all. This interference method was first realized in 1965 by U. Bonse and M. Hart for X-rays \cite{Bonse1965, BonseHart1965} and nine years later, in 1974, for neutrons \cite{RauchTreimerBonse}.  This behavior is fundamentally different from a light interferometer of the same type (Mach Zehnder).  

First, the interference pattern in front of or inside the third crystal plate (analyzer crystal) contains all information transmitted by the two coherent partial waves along path I and path II, e.g.  crystal defects in the interferometer, phase changes  and any  misalignment. 

\begin{figure}[h]
 \centering
 \textbf{(a)} \includegraphics[width=0.9\columnwidth]{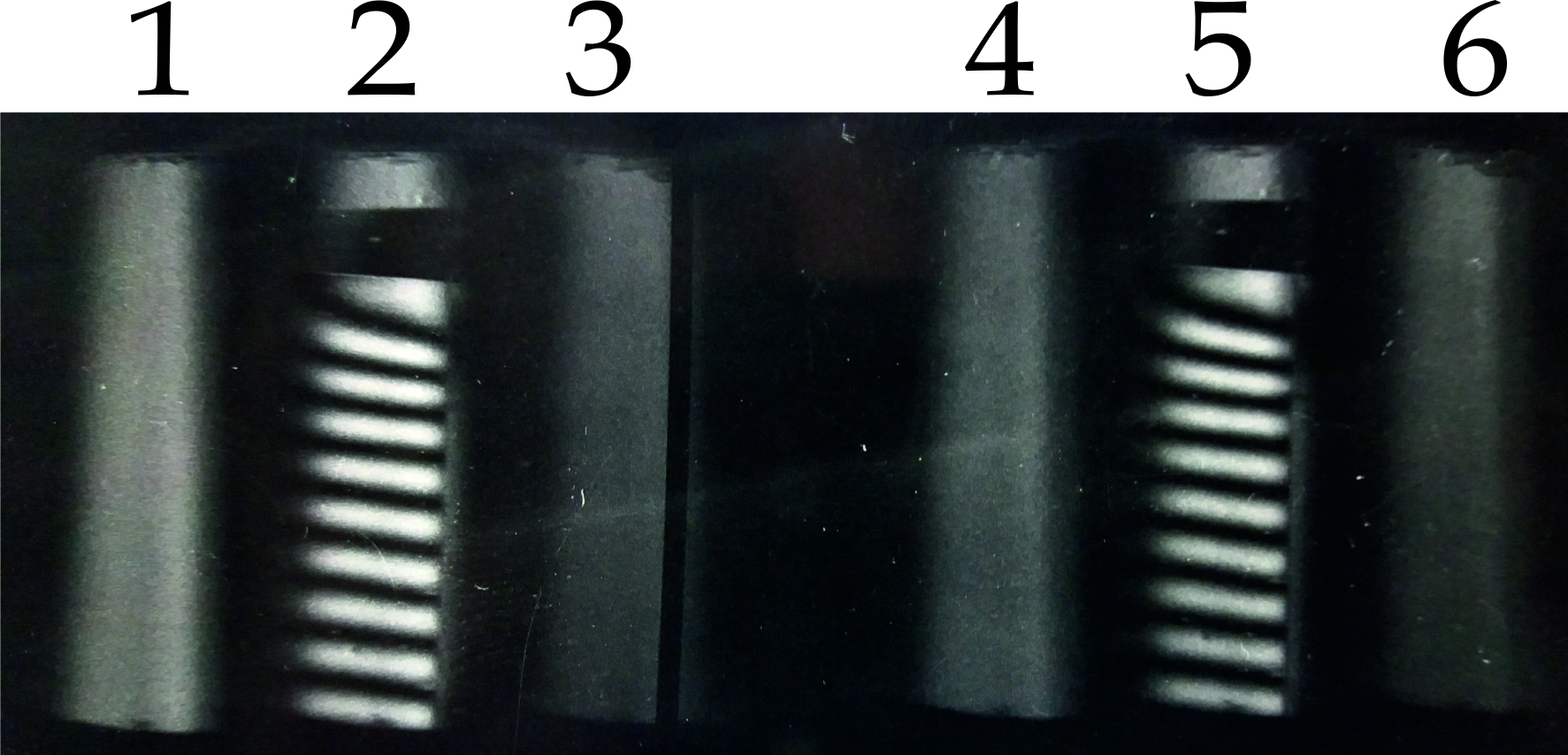} \\ 
\textbf{(b)} \includegraphics[width=0.9\columnwidth]{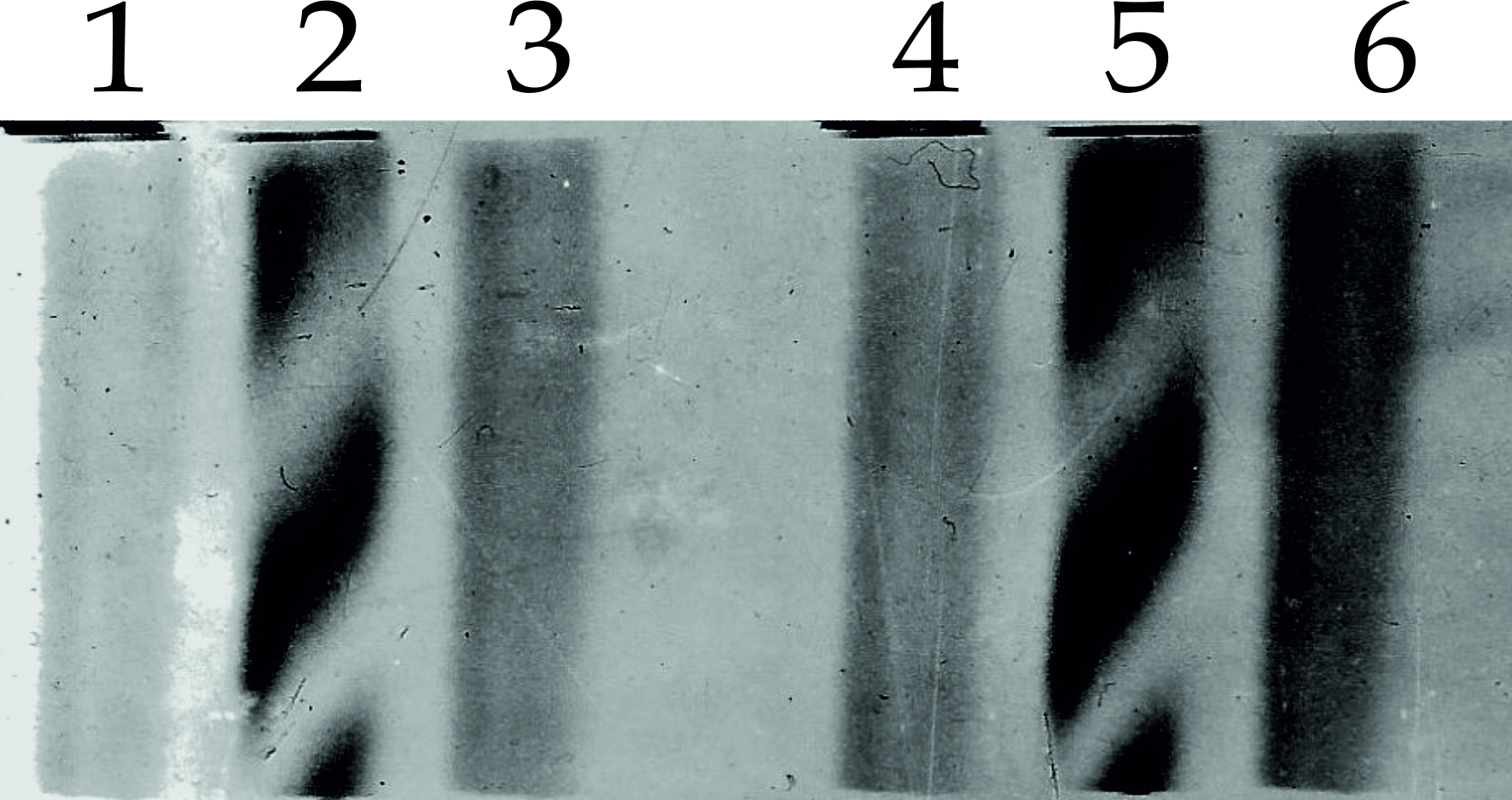} 
 \caption{\textbf{Two different X-ray interferograms (a) and (b)}  $|$  Each interferogram  has two pairs of three fringes (because all rays could  pass the crystal interferometer), no 2 and no 5 are the interference patterns  in the O - and  G - direction.  
Upper image:  A rotation of analyzer crystal (third crystal plate) about 10$^{-8}$ rad against the interference pattern  produced a so-called rotational Moiré pattern.  Lower  image: A vertical  Beryllium wedge in one beam path  in the interference beam produced horizontal fringe pattern. (Mo K$ \alpha_{1,2}$, exposure time = 3h, width of each pattern $ \sim $4mm ). Note, that X-rays could penetrate all three Si crystal plates, each 4.460 mm thick, only because of the anomalous absorption,  the  so-called "Borrmann effect"  \cite{Borrmann1950};  in the case of neutrons (no absorption), the entire "Borrmann Delta" contributes to neutron interference, but only one "pair of coherent rays" at a time  (original images are from \cite{TreimerThesis}). }
 \label{Rotmoiré}
 \end{figure}

Second, the Moiré amplification mentioned above works only  in the [x] -  direction (see Fig.~\ref{Fig_1}) \cite{Bonse1965}.  
This was clearly demonstrated in test experiments of the neutron interferometer with X-rays (1973), see \cite{TreimerThesis}) by generating so-called rotational Moiré patterns (upper image in Fig.~\ref{Rotmoiré}) and phase shifting by a vertical Beryllium wedge in the [z] - direction (lower image in Fig.~\ref{Rotmoiré}).

Third, in the case of  X-rays only one wave field  is excited, traversing the crystal parallel  to the reflecting atomic planes,  the other {{wave field }}  is absorbed. Due to  an  anomalous low  absorption \cite{Borrmann1950},  X-rays can pass also  "thick" perfect crystals.  In an X-ray interferometer, the O-beam and the G-beam have the same intensity distribution.

\begin{figure}[h]
\centering
\includegraphics[height=3.9cm]{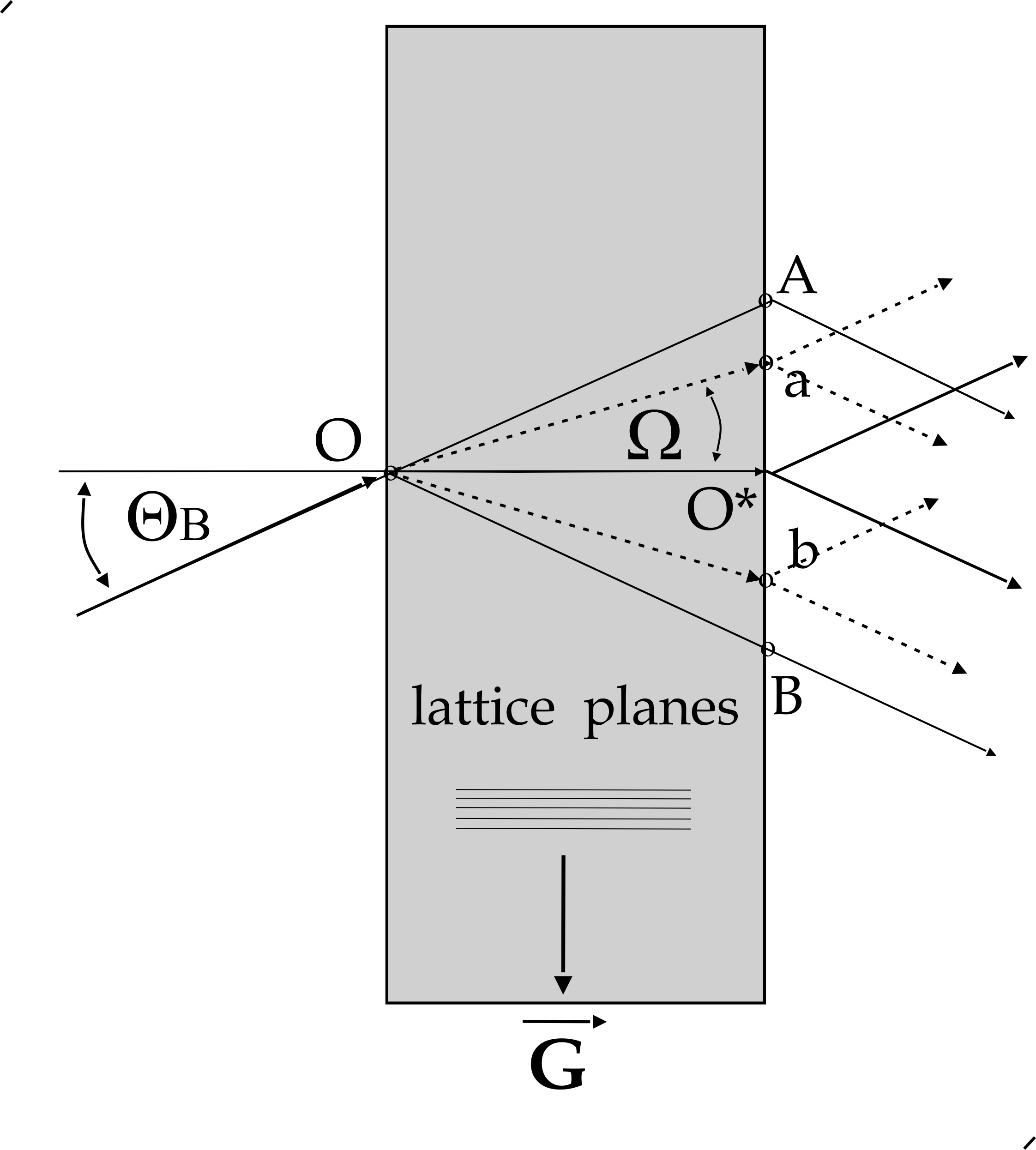} 
\includegraphics[height=3.9cm]{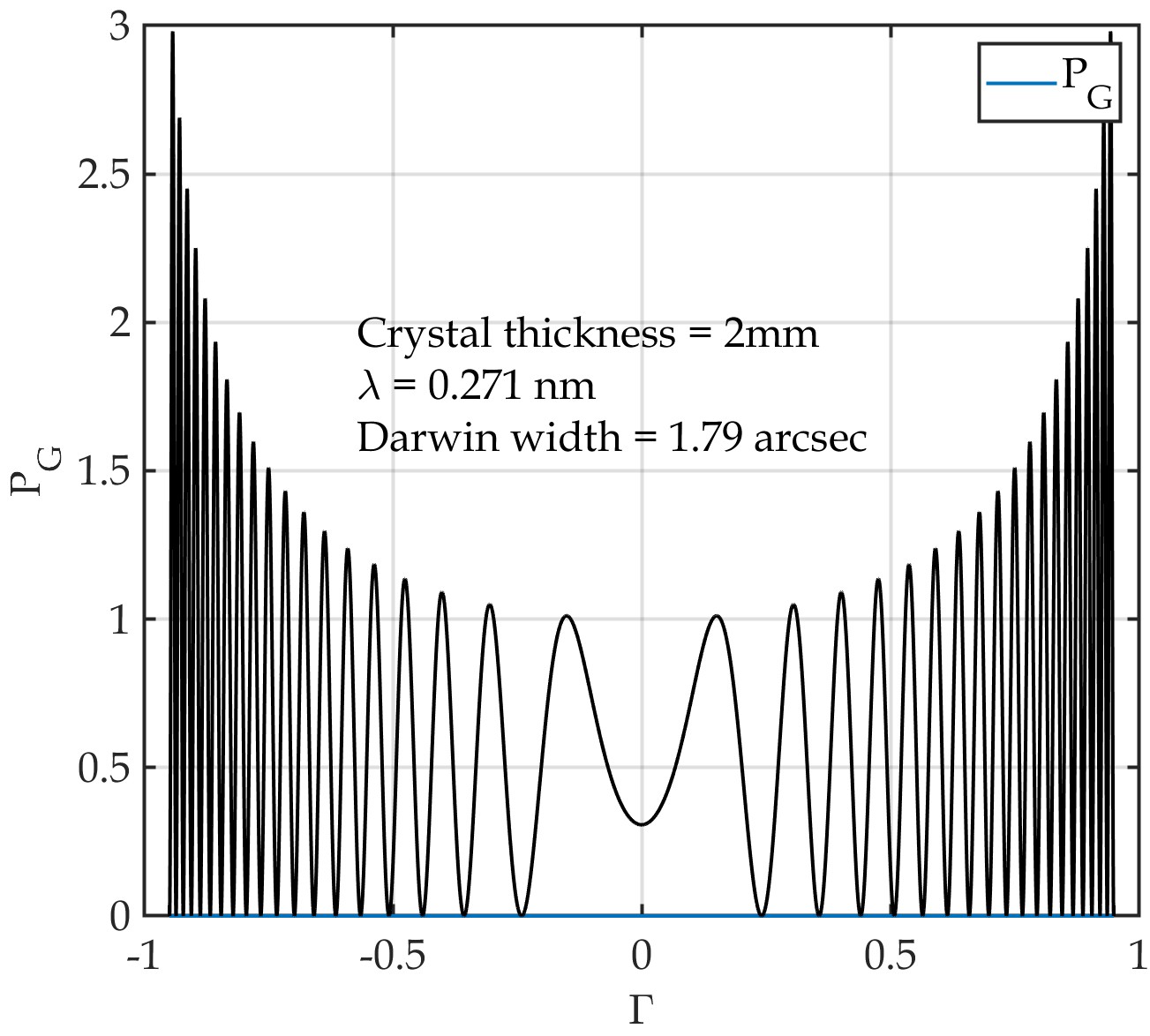} 
\caption{\textbf{Laue diffraction:  Rays and intensity distribution}  $|$ Left  image: Rays in the crystal  in the case of  so-called Laue diffraction.  Triangle $\widehat{OAB}$ = "Borrmann Delta",  for an incident pencil beam at  Bragg angle ${\Theta _B}$.   'a' and 'b' are the exit points of waves coming from the $ \alpha $ and $ \beta $ branch  of the dispersion surface.   Right image:  Laue diffracted intensity distribution behind a crystal plate,  $\Gamma  = \tan (\Omega )/\tan ({\Theta _B})$. \cite{RauchWerner2000} }
\label{Borrmann}
\end{figure}

For neutrons with a  wave length $ \lambda$, and  $ \Delta\lambda/\lambda\ll$10$^{-4}$ always the whole "Borrmann Delta" is excited (see Fig.~\ref{Borrmann}) because for any other  neutron {{wave}} incident at  exactly the same direction but different wave length, the energy flow is split within the "Borrmann Delta" and  exits  at "a" ($ O^{*} < a < A$)  and "b", ($ O^{*} < b < B $)  (see left image in Fig.~\ref{Borrmann}).  This is also true for neutrons whose angle of incidence in the $ \{x,y \}$ plane deviates from the Bragg angle by less than 10$ ^{-7} $ rad \cite{Petrascheck1988,Shull1968}  (see right  image in Fig.~\ref{Borrmann}). 
Therefore, interference can only occur due to "one-to-one interference" between the wave from path I and the wave from path II (see above Fig.~\ref{Fig_1})  and only within their scattering plane ("pair of coherent beams"  \cite{BonseHart1965}).  Note, that there is only one neutron in the interferometer at a time and only self-interference occurs.

All information contained in the interference pattern of such a crystal  interferometer is therefore  amplified one-dimensionally, only.
Moreover,  due to the extremely large difference  of the lateral coherence lengths $\sigma_x $ and $\sigma_z$ and due to the rather complicated intensity structure resulting from the "Borrmann Delta", observable interferences outside the scattering plane ($ \{x,y \}$-plane) are extremely unlikely. 
The neutron wave is  laterally limited and can be described by a Gaussian like  wave packet but with different  lateral coherence lengths $\sigma_x $ and $\sigma_z$.  In \cite{Clark} $\sigma_{z} $ and $\sigma _{x} $ were estimated as  60nm and a few $\mu $m, respectively.  
Therefore, any neutron interference process takes place only within an extremely thin layer  ($ \ll 1\mu $m) of the scattering plane by a strong one-to-one correlation between rays of path I and path II  (see  \cite{RauchSuda1974,PetrascheckRauch1984,Petrascheck1988}).

Concerning a  SPP-induced neutron OAM (n-OAM), experiments with laser light using a  SPP have shown that incident wave packets must be able to detect the discontinuity at the center of the vortex phase plate, otherwise the probability of detecting a single particle with uniform OAM around the center of the packet is extremely low \cite{Cappelletti2020}, which is certainly  true for neutrons in a  crystal  interferometer,  and even for X-rays  it is  difficult to use a division amplitude interferometer (like the used neutron interferometer) to create a vortex wave front with SPP  \cite{Peele2004}.

In order to prove that no n-OAM is necessary to arrive at the experimental results of \cite{Clark},  we calculated all the interference patterns of  their experiments with exactly the same parameters, i.e., neutron wavelength,  SPP material,   Si crystal neutron interferometer.  
We assumed  a perfect crystal  with the shape as shown in Fig.~\ref{Fig_1} and calculated each interferogram  given \cite{Clark} on the basis of dynamical neutron diffraction, only  \cite{RauchSuda1974,PetrascheckRauch1984}.  

The spiral phase plates (SPP) in \cite{Clark} had different sizes (10mm and 15mm diameter), we used a 15mm diameter SPP with a thickness height of one and two $ \lambda $-thicknesses D$ _{\lambda} $  as in \cite{Clark} (see Section~\ref{Theory}  for all calculations). 
Each of the neutron experiments, as described in [1], lasted more than three days due to the low neutron count rate, therefore,  despite data processing,  the counting rate did not appear to be uniformly distributed over the entire detector area, perhaps due to inhomogeneity of the neutron beam, perhaps due to varying efficiency of individual detector pixels, perhaps due to small changes in the phase stability of the interferometer.  This can be seen in all interferograms  in \cite{Clark}, best in the figure "Extended data Figure 1 | Raw data".  These  deficiencies were  not simulated. 


Here the experimental interferograms taken from \cite{Clark} are presented (left parts) together with the corresponding interferograms obtained using dynamical theory of neutron diffraction.  The mathematics is given below. \\

\section{Calculation of interferograms}\label{Theory}

\vspace{0.5cm}
In a Mach-Zehnder light interferometer, the coherence of  beam is defined in front of the interferometer, while in a crystal interferometer of Mach-Zehnder type the interference property is directly related to the Laue-Bragg reflections (of neutrons and  X-rays) in the interferometer.
In both interferometers,  the phase of one partial wave can be  manipulated with respect to the other partial wave with a phase-shifting object that has a so-called $ \lambda $-thickness ${D_\lambda } $,  for neutrons  ${D_\lambda } = 2\pi /( N {b_c} \lambda ) $, ($ b_{c} $ is the neutron coherent scattering length, N is the number of atoms/unit volume, $\lambda$ =  0.271nm  is the neutron wave length). ${D_\lambda } $ is the thickness of a material that shifts the phase of one partial wave relative to the phase of the other partial wave by 2$ \pi $. 

In \cite{Clark} an "effective angular momentum"  is given as  $ L = N  {b_c}  \lambda   {h_s}/2\pi $,  which can be rewritten as $ L = {h_s}/{D_\lambda }$,  h$ _{s} $ being  the step height of the spiral.  
  This means that L in this context is always a linear function of step height h$ _{s} $ scaled by ${D_\lambda } $ and therefore has no quantization  l$\hbar$. 
If L is allowed to take only integer values (1, 2, ... n), i.e. a multiple of ${D_\lambda } $,  as done in \cite{Clark}, then one can only measure interferograms as given in \cite{Clark}, and thus apparently observe a quantization of L,  but  L can take any real value $ \in R $. 
In our calculations, for example, the value L = 7.5 as in \cite{Clark},  simply leads to the same interferogram as in Fig.~\ref{L1234}, without assuming any quantization of L. 

\begin{figure}[hbt]
\centering
\includegraphics[height=2.5cm]{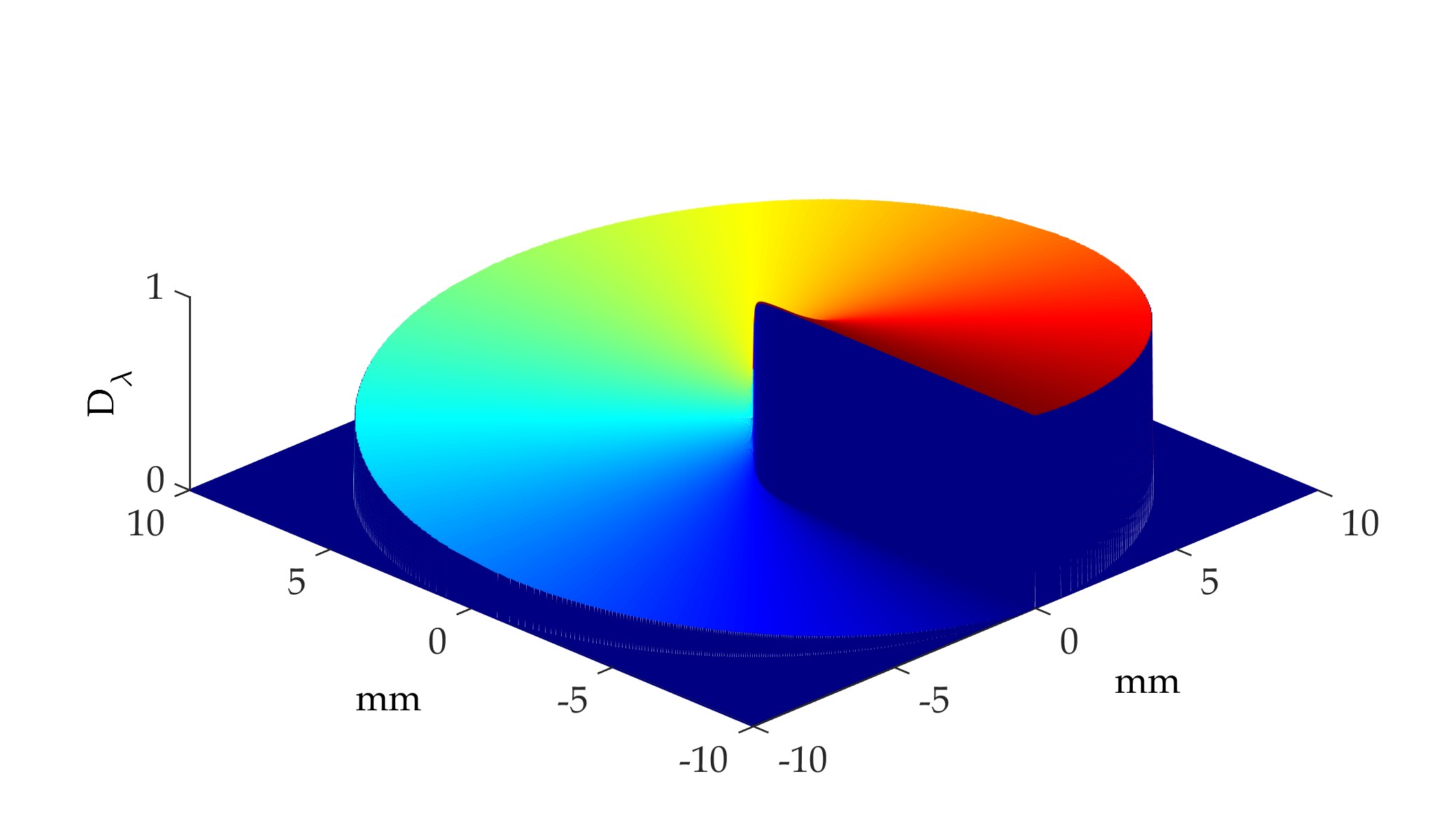}   
\includegraphics[height=2.5cm]{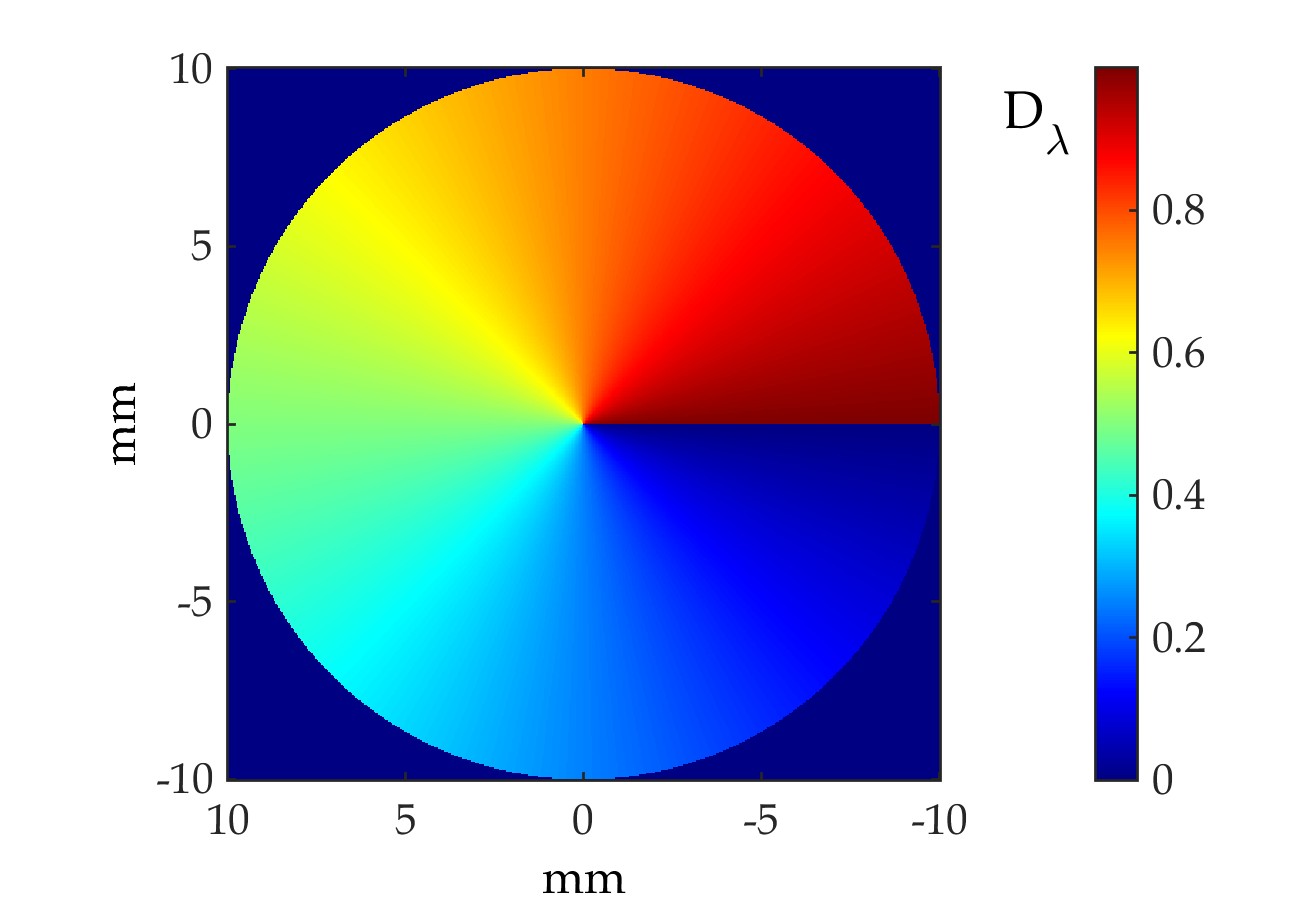} 
\caption{\textbf{Spiral phase plate (SPP) and its Radon transform}  $|$ Left image:  3D image of the Aluminum spiral phase plate, step height   $D_{\lambda} $  = 112$ \mu $m. Right  image:  The Radon transform of the SPP, scaled by  $D_{\lambda} $  of the SPP.} 
\label{spiralphase}
\end{figure}

We first calculated the 2-dimensional  phase front of the spiral phase plate (SPP) using the Radon transform for only one incidence angle \cite{Radon1917}, $RT\{SPP\} $, which provides the spatially dependent transmission lengths of a beam for a scan angle $ \alpha $ (Eq.~\ref{RT}).  
The SPP was divided into N z-slices and the corresponding $RT\{SPP_{z}\} $ calculated as 
\begin{eqnarray}
\begin{array}{l}
{{\rm{RT}}_{0,\alpha }}\{ SPP_z\}  =  \int\limits_{ - \infty }^\infty  {\int\limits_{ - \infty }^\infty  {SP{P_z}(x,y) \cdot \hat \delta  \cdot dx \cdot dy} } \\
{\rm{with}}\quad \hat \delta  = \delta [p - x \cdot \cos (\alpha ) - y \cdot \sin (\alpha )]
\end{array}
\label{RT}
\end{eqnarray}
Subscript  0  stands for the orientation of the  SPP in a coordinate system $ \{x,y,z\} $,  $ \alpha $  is  the scanning angle of the Radon transform (here $\alpha = 0$,  i.e. only one projection is used),  p is the sampling parameter for parallel scanning of the zth slice,  subscript z  denotes  the z-th slice of the SPP. 
Finally, the $RT\{SPP\} $ was scaled by the refractive index of the SPP material (Al), resulting in a 2D phase front behind the SPP (see Fig.~\ref{spiralphase}).  The left  image in Fig.~\ref{spiralphase} shows the spiral phase plate, step height is $ D_{\lambda} $ =  112$\mu m$ , the right  image  shows the 2D Radon Transform  of the  SPP, scaled by $ D_{\lambda} $ of the SPP.  
This was verified by a different computation, where the 2D thickness function of the SPP (for a scanning angle $\alpha = 0$) was computed  directly from the geometry of the SPP   as a phase-shifting object.  


%
For the calculation of the interference pattern we used the same formula as used in neutron interferometry \cite{RauchWerner2000} and thus also by \cite{Clark}.
\begin{equation}
I(u,v) \propto {\rm{ }} \cos{\rm{\big(}}{\phi _0} + 2\pi  \cdot {D_S}(u,v)\big)\
\label{Equ:Iuv}
\end{equation}
Here I(u,v) denotes the two-dimensional intensity distribution that would be measured  by a position sensitive neutron detector, $ \phi_o $ is the phase of the partial wave, which can be tuned with a phase flag (as was used  in \cite{Clark},  Fig.~\ref{Fig_1}), $ D_{S}(u,v)$  is the position  dependent thickness, at the point, where the neutron wave transmits the SPP, scaled by  $D_{\lambda} $.
In order to compare with the experimental interference pattern of \cite{Clark},  we used the resolution of the neutron detector as discretization parameter in the $x$-$z$ - plane ($ 100 \times 100$) and added some random noise. All calculations have been performed with MATLAB~\cite{Matlab}.   

 The results  of our calculations are shown in Figs.~\ref{L1234} - \ref{L3phi_var}, where we compare the experimental results of \cite{Clark} with our simulated interference pattern.
In Fig.~\ref{L3phi_var} we also display the effective $ \Delta $D due to rotation $\phi_0$ of the phase flag. 
It can be seen that for these small rotations,  the $\Delta $Ds  are  only a fraction of   $D_{\lambda} $ = 112$\mu m $  and thus a noticeable phase shift is hardly observable (see Fig.\ref{L3phi_var}), i.e. one cannot expect a dramatic change of the interference patterns due to such small phase shifts, and also due to the low  counting rate of neutrons. 

Concerning Fig.~4,  in \cite{Clark} one reads:  {"The average OAM is independent of the position of the phase flag, i.e. L is preserved"}. This is a thoroughly misleading interpretation, since  similar looking interferograms do not prove  L conservation.
In fact, the number of neutrons is always preserved if the phase in  one  partial beam in a neutron interferometer is changed, well known as conservation of the number of particles in neutron interferometry.
 
 Furthermore, due to the small phase shifts, one cannot in any case conclude that L is conserved, because  without knowing the counting  rate of the O-detector (see Fig.~\ref{Fig_1}), one can by no means say unequivocally that anything is conserved.
One has to question why only such small flag angles $ \phi_0 $ were used to demonstrate a  so-called "conservation of L", which should actually hold for all flag angles $ \phi_0 $.  With  much larger flag angles, they would have observed with the O and G detectors  only a rotation of the whole interference pattern, which again would only prove the conservation of the neutron number and contradicts the assumption of partial prism refraction as the cause of an n-OAM.
In neutron interferometry, the O-detector always counts the complementary number of neutrons of the G-detector, so that only the {sum} of the two counting rates  can tell us anything about the conservation of a quantity. 
The O-detector in \cite{Clark} was (unfortunately) used as an "integrating counter", probably to monitor the low neutron flux. Real monitoring is usually done with a transmission detector in front of the interferometer. 
If Clark et al. had also used the "O detector" for complete neutron counting  and not as a monitor (see Fig.~\ref{Fig_1} in \cite{Clark}), they would have observed complementary neutron interferograms to those of the G detector and could have inferred only conservation of particle number.

One  important feature of quantized OAM is the superposition of different, opposite orbital angular momentum, $\left| { + l} \right\rangle  + \left| { - l} \right\rangle$ resulting in a ring-like structure,  containing exactly 2$ \cdot $l intensity maxima, so  the far more interesting neutron experiment, and thus demonstration of the controlling  of n-OAM, would have been to image (with no crystal interferometer) the coherent sum of two opposite spiral structures leading to opposite OAMs, as was done in \cite{Fickler2018}.\\

\newpage

\begin{widetext}

\section{Results}

\begin{figure}[h!]
\includegraphics[height=4.5cm]{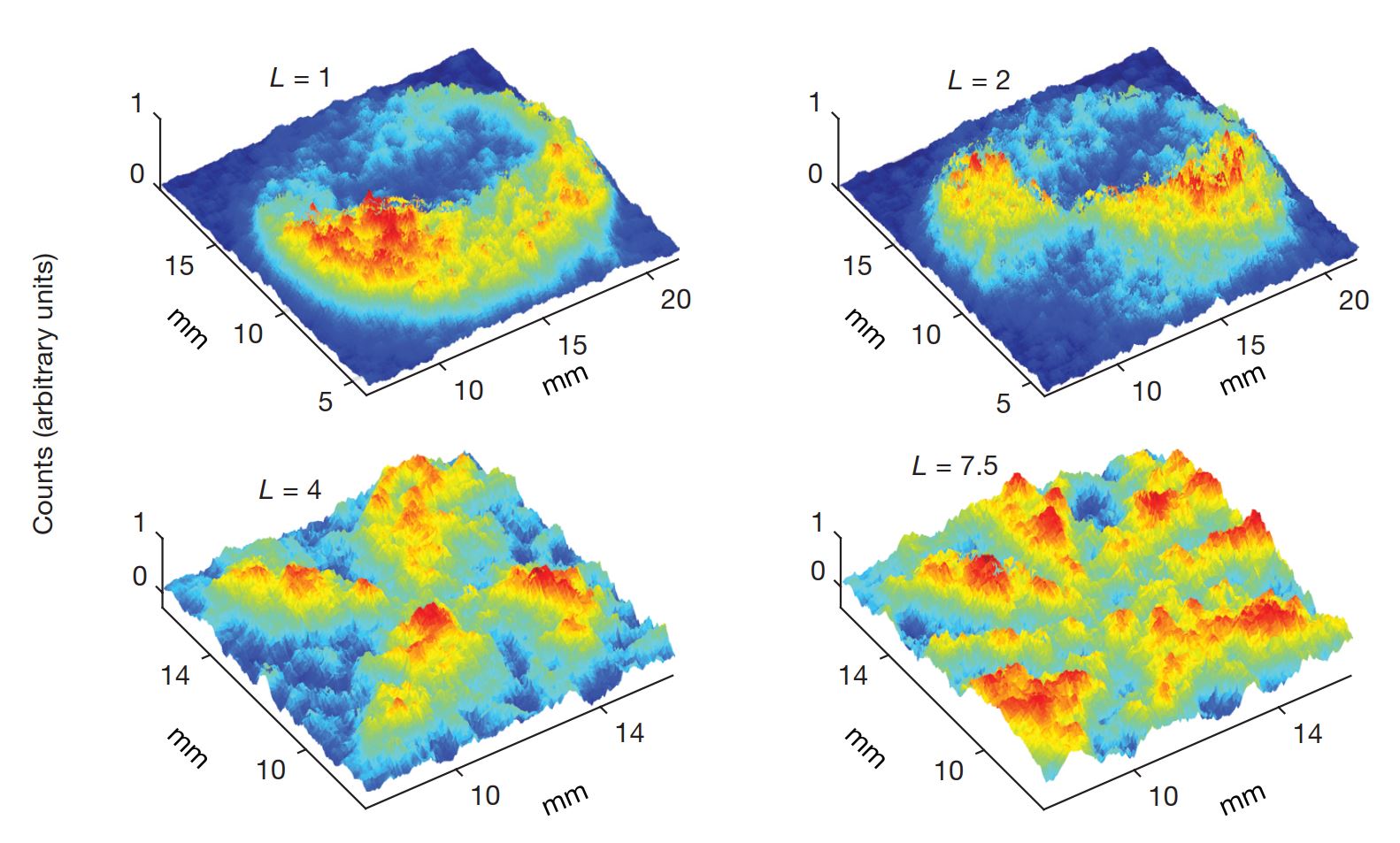} \hfill\vline\hfill
\includegraphics[width=7cm]{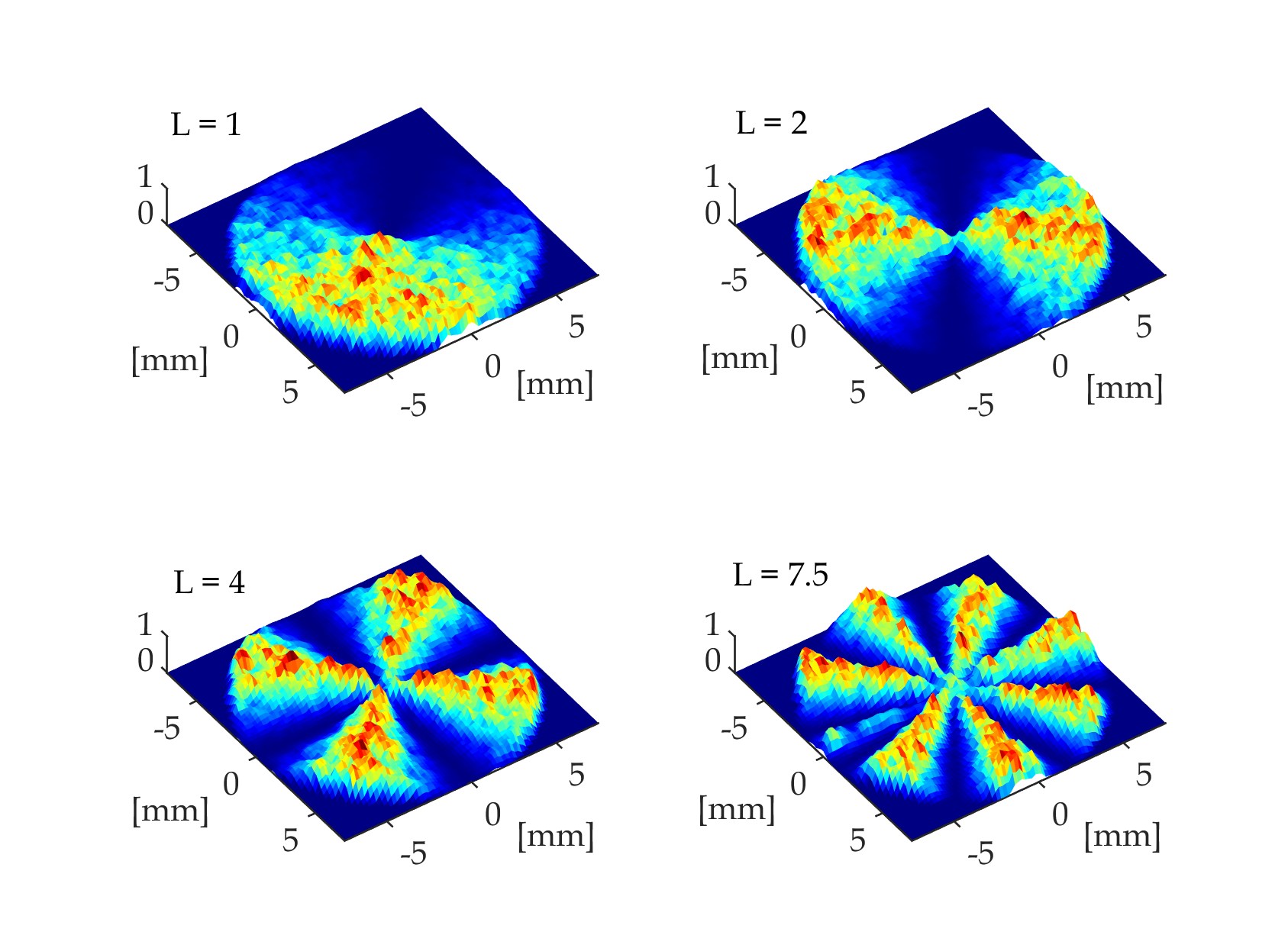} 
\caption{\textbf{Comparison of  interferograms} $|$
Left part: Interference patterns  showing "so-called" n-OAM  for L =1 , L = 2, etc. presented in \cite{Clark},  Figure 2.  Right part: Interference patterns simulated with dynamical neutron diffraction without assuming n-OAM.}
\label{L1234}
\end{figure}
\begin{figure}[h!]
\includegraphics[height=5.9cm]{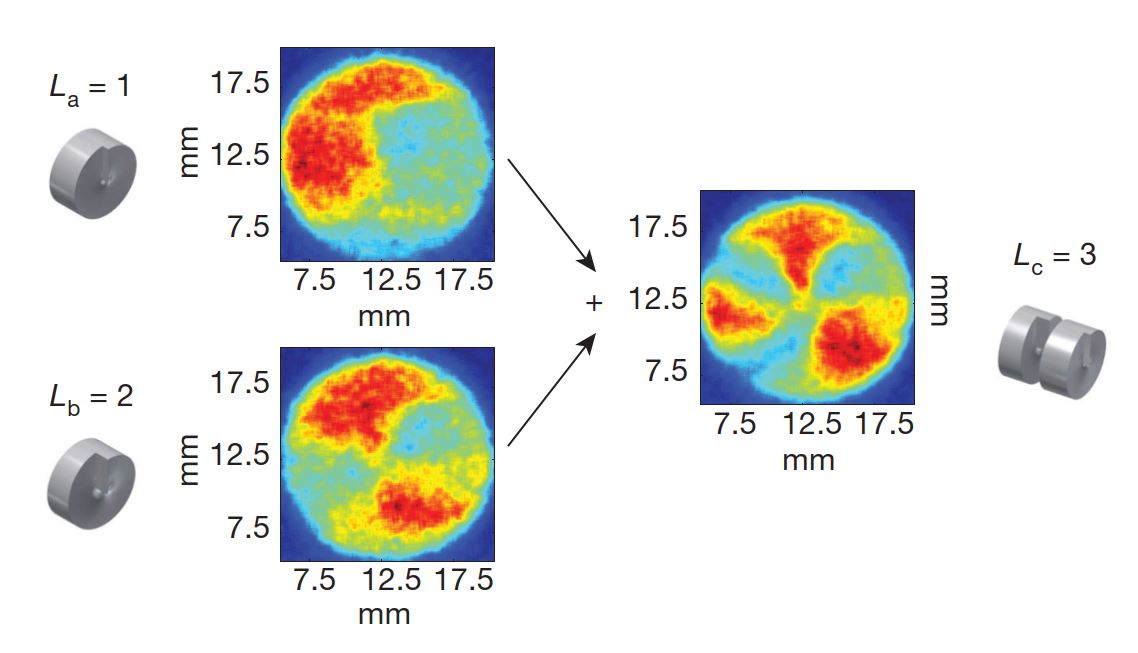}  \vline \hspace{0.2cm} \hfill 
\includegraphics[height=5.9cm]{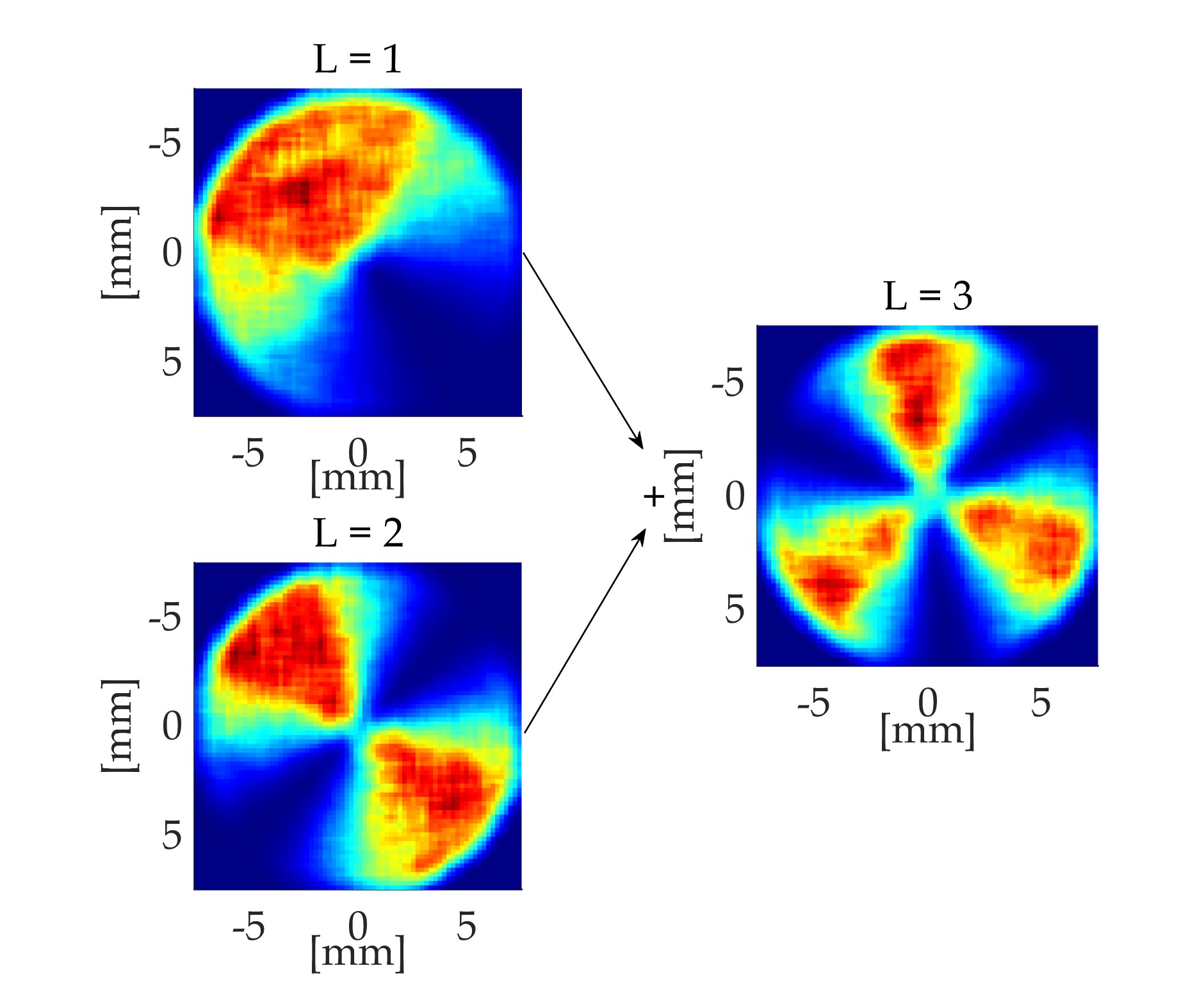} 
\caption{\textbf{"So-called" addition of n-OAM}  $|$
Left part: Experimental interference patterns taken from \cite{Clark}, Fig 3. Right part:  Interference patterns simulated with dynamical neutron diffraction without assuming n-OAM. }
\label{L1plus2=3}
\end{figure}
\begin{figure}[h!]
\includegraphics[height=3.cm]{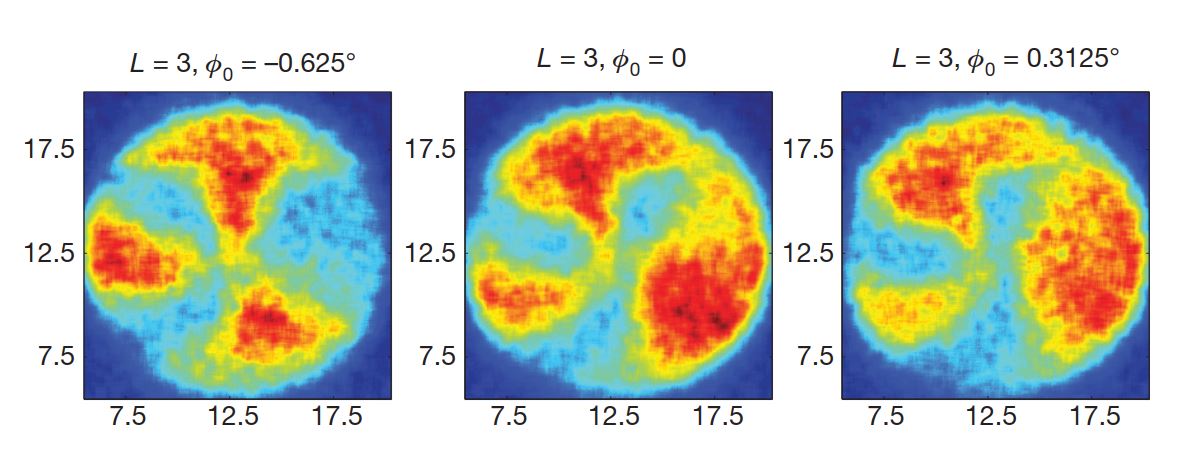}   \hfill \vline 
\includegraphics[height=3.2cm]{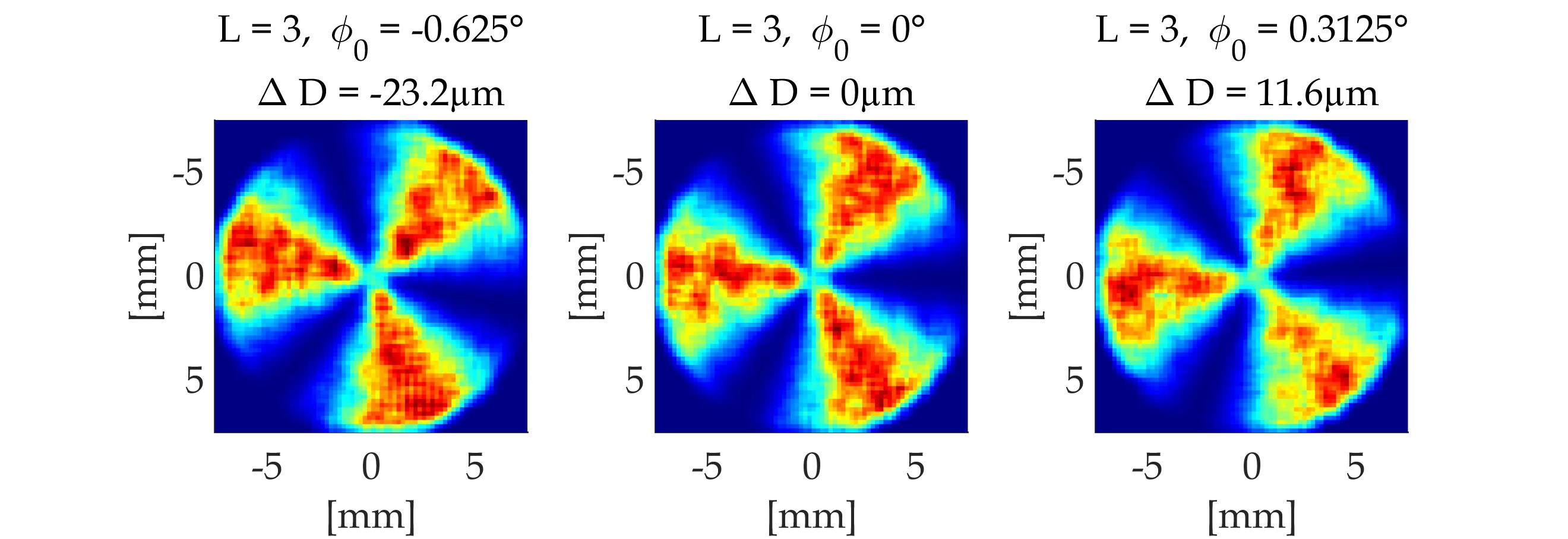} 
\caption{\textbf{"So-called" rotational invariance}  $|$
Left part: Experimental Interference patterns taken from \cite{Clark}, Fig 4.  Right part: Interference patterns simulated with dynamical neutron diffraction without assuming n-OAM. Note that  the resulting change of  phase plate thickness $ \Delta $D  due to a  rotation of $ \phi_o $  is only a small fraction of  $D_{\lambda }$ = 112$\mu m $.}
\label{L3phi_var} 
\end{figure}
\vspace{1.5cm} 
\end{widetext}

\newpage
\section{Conlusion}

In summary, the interferograms we calculated are in perfect agreement with  the interferograms in the Clark publication \cite{Clark}  and showed that there is definitely no reason to interpret the interferograms in the context of a  n-OAM.
Moreover, here are several serious reasons that Clark et al.  could not have measured  a n-OAM. 
One important reason is that the lateral coherence is too small,  which was already proved  in \cite{Cappelletti2018} and in \cite{Jach2022}.  
The arguments given there  are correct, but incomplete,  because the coherence of the partial waves  in the interferometer  has {two} laterally very different  coherence lengths, that are both  too small and moreover  extremely different in size, as already stated above.  
Any assumption of  transverse wavefunction  of a neutron wave packet as in \cite{Sarenac2016} and in other publications, as 
\begin{equation}
{\Psi _t} = \frac{1}{{\sqrt {2\pi {\sigma^2}} }}\exp \left[- {\frac{{{{(x - {x_0})}^2} + {{(z - {z_0})}^2}}}{{{4}\sigma^2 }}} \right]
\end{equation}
(our notation is x,z),  assuming $ {\sigma _x} = {\sigma _z} \equiv \sigma $ the coherence length,  ignores that  $ {\sigma _z} \ll {\sigma _x} $,  given also in \cite{Clark} and  measured in \cite{Pushin208}. 
{{Thus, we agree with  A. Luski et al. \cite{Luski2021}: "There have been experimental efforts to create neutron beams with OAM (25) = \cite{Clark}, which have been hindered by insufficient coherence lengths (26)=\cite{Cappelletti2018}. Thus far, no experiment has created vortex beams of nonelementary particles".((25) is Luski's reference to our reference [1], (26) corresponds here to [16]).} }

{{An  argument that the n-OAM consists of a sum of refracting radial prisms, which in sum yields an observable n-OAM, can be easily falsified by Fig.~\ref{L1234} and Fig.~\ref{L1plus2=3}. In both Figures, the phase shift is increased by additional SPPs in one beam path of the interferometer.  If refraction were the primary interaction (as assumed above), then the interference pattern would  not show additional regions of scattered neutrons, as in  Fig.~\ref{L1234} and Fig.~\ref{L1plus2=3}. 
Adding one, two, or more SPPs in one  beam path, as is done in \cite{Clark}, causes pure phase shifts, that produce these interferograms, not refraction.  
A simple calculation of the change in direction of the neutron wave due to an average SPP refraction yields a deviation of about  $ 4 \times 10^{-9} $ rad, which is much too small to produce a (twist-based) n-OAM.}
}

The most serious reason for the misinterpretation of the neutron interferograms published by Clark et al. is ignoring the completely different operation of a Mach-Zehnder light interferometer and a crystal neutron interferometer (of Mach-Zehnder type), which have the same  topology  but function differently.
Both are amplitude splitting interferometers, but in contrast to a Mach-Zehnder light  interferometer, in a neutron interferometer each phase change caused  by a phase-shifting material  can only be detected via Moiré pattern. Only the superposition of the interference pattern (as a standing wave) with the reflecting crystal planes of the third crystal plate of the neutron interferometer makes the observation possible, but the amplification works only in one lateral direction (in this paper in  x-direction, see Fig.~\ref{Fig_1}). 
This is in contrast to a Mach-Zehnder light  interferometer, where the interference is solely caused by superposition of the two partial waves.
  
The large differences in wavelength, refractive index and  coherence properties of neutrons as  compared to photons,  the different modes of reflection, the extreme different shape of  neutron Gaussian wave packets underline  the big doubts that the detection of a n-OAM by means of a neutron crystal interferometer as used in \cite{Clark} is possible at all.  We conclude,  that all interferograms in \cite{Clark} are pure phase contrast images of the spiral phase plate and that no control of neutron orbital momentum has been detected in \cite{Clark}. {{This is also true for other experiments where one wants to detect or even use n-OAM with a crystal neutron interferometer.
}
}
However, there is no doubt about the existence of neutron angular momentum, but its detection must  be done in a very different, unambiguous way.  \\

\textbf{Acknowledgments} We thank  G. Badurek,  T. Jach and P. Schattschneider for helpful comments and valuable discussions, especially D.  Petrascheck, whose critical comments and advice contributed a lot to the success of this publication.
One of us (W.T.) thanks the Helmholtz Zentrum Berlin  for its hospitality. \\

\textbf{Author Contributions} W.T. had the idea,  F.H. calculated the spiral phase plate, W.T. and F.H.  calculated the interference patterns,  M.S. contributed the mathematics  concerning dynamical theory, neutron interferometry, etc.   W.T.,  F.H. and  M.S. wrote the manuscript. \\

\textbf{Author Information}  The authors declare no competing financial interests.\\

\textbf{Code availability statement} We used MATLAB~\cite{Matlab}   to calculate the interferograms  presented in Fig.~5-7. The MATLAB-scripts  can be accessed via a public Gitlab-repository  \url{https://gitlab.com/noam7055430/spp\_interferograms} \\

\textbf{Correspondence}  Correspondence and requests for materials should be addressed to W.T. \\ 
 (wolfgang.treimer@bht-berlin.de).\\

---------------------------------------------------------------

\end{document}